%% file: main.tex
\documentclass{sigchi}

\CopyrightYear{2020}
\setcopyright{acmcopyright}
\doi{https://doi.org/10.1145/3313831.3376726}
\isbn{978-1-4503-6708-0/20/04}
\conferenceinfo{CHI'20,}{April  25--30, 2020, Honolulu, HI, USA}
\acmPrice{\$15.00}

\usepackage{balance}       
\usepackage{graphics}      
\usepackage{graphicx}
\usepackage{subcaption}
\usepackage{txfonts}
\usepackage{mathptmx}
\usepackage[pdflang={en-US},pdftex]{hyperref}
\usepackage{color}
\usepackage{booktabs}
\usepackage{textcomp}

\usepackage{multirow}

\usepackage{mfirstuc}
\usepackage{paralist}

\usepackage{fontawesome}

\usepackage[dvipsnames]{xcolor}
\definecolor{fast}{HTML}{ED6940}
\definecolor{slow}{HTML}{FF903E}
\definecolor{louder}{HTML}{84B13D}
\definecolor{softer}{HTML}{B7CD49}
\definecolor{micropause}{HTML}{6FB1FC}
\definecolor{masterpause}{HTML}{4387ED}
\definecolor{longpause}{HTML}{0052D4}
\definecolor{stress}{HTML}{9A86D6}
\definecolor{normal}{HTML}{A2A2AC}

\usepackage{microtype}        
\usepackage{ccicons}          

\usepackage{todonotes}

\def\plaintitle{SIGCHI Conference Proceedings Format}

\def\emptyauthor{}
  \def\plainkeywords{Voice modulation; evidence-based training; data visualization; public speaking.}

\makeatletter
\def\url@leostyle{%
  \@ifundefined{selectfont}{
    \def\UrlFont{\sf}
  }{
    \def\UrlFont{\small\bf\ttfamily}
  }}
\makeatother
\def\pprw{8.5in}
\def\pprh{11in}

\definecolor{linkColor}{RGB}{6,125,233}
\hypersetup{%
  pdftitle={\plaintitle},
  pdfauthor={\emptyauthor},
  pdfkeywords={\plainkeywords},
  pdfdisplaydoctitle=true, 
  bookmarksnumbered,
  pdfstartview={FitH},
  colorlinks,
  citecolor=black,
  filecolor=black,
  linkcolor=black,
  urlcolor=linkColor,
  breaklinks=true,
  hypertexnames=false
}

\usepackage{cuted}
\usepackage{capt-of}

\newcommand{\name}{{\textit{VoiceCoach}}}
\newcommand{\up}{{user panel}}
\newcommand{\rv}{{recommendation view}}
\newcommand{\pv}{{practice view}}
\newcommand{\vtv}{{voice technique view}}

\begin{document}

\title{VoiceCoach: Interactive Evidence-based Training for Voice Modulation Skills in Public Speaking}

\numberofauthors{1}

\author{%
  \alignauthor{Xingbo Wang, Haipeng Zeng, Yong Wang\thanks{Corresponding author.}~, Aoyu Wu, Zhida Sun, Xiaojuan Ma, Huamin Qu\\
    \affaddr{Department of Computer Science and Engineering, \\The Hong Kong University of Science and Technology, Hong Kong, China}\\
    \email{\{xingbo.wang, hzengac, ywangct, awuac, zhida.sun\}@connect.ust.hk, \{mxj, huamin\}@cse.ust.hk} }\\
\vspace{-6mm}
}

\maketitle

\begin{strip}\centering
\includegraphics[width=0.9\textwidth]{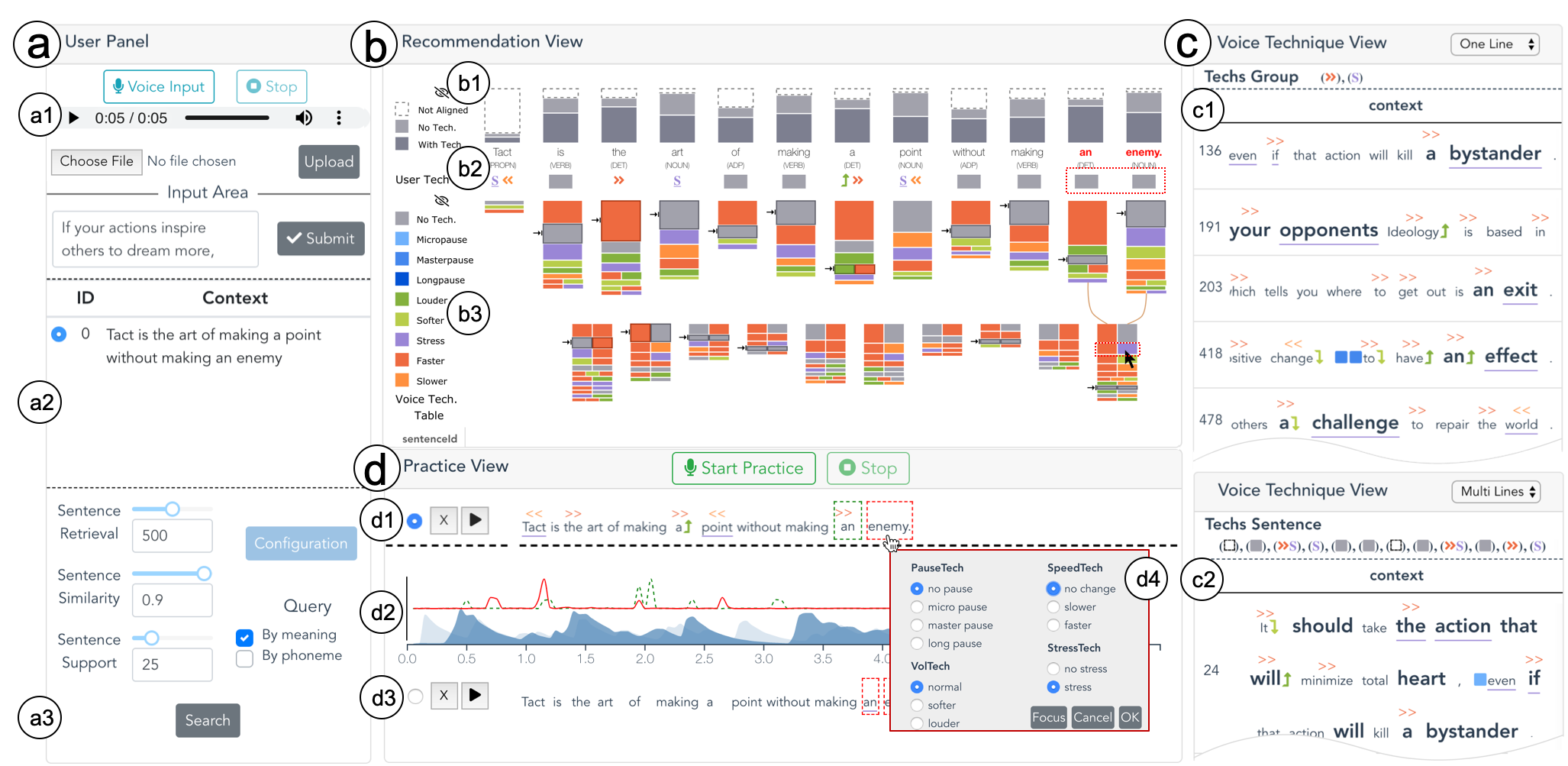}
\vspace{-3mm}
\captionof{figure}{\textcolor{black}{The user interface of {\name}: (a) The {\up} allows users to submit a query sentence via audio or text input. (b) The {\rv} presents different levels of recommendation results of modulation combination. (c) The {\vtv} enables users to quickly locate and compare the contexts of a specific voice modulation skill in either one-line mode or multi-line mode. (d) The {\pv} provides users with real-time and quantitative visual feedback to
iteratively practice voice modulation skills.}}
\label{fig:systemInterface}
\end{strip}


\begin{abstract}
The modulation of voice properties, such as pitch, volume, and speed, is crucial for delivering a successful public speech. 
However, it is challenging to master different voice modulation skills.
Though many guidelines are available, they are often not practical enough to be applied in different public speaking \textcolor{black}{situations}, especially for novice speakers.
We present \textit{VoiceCoach}, an interactive evidence-based approach to facilitate the effective training of voice modulation skills.
Specifically, we \textcolor{black}{have analyzed} the voice modulation skills from 2623 high-quality speeches (i.e., \textit{TED Talks}) and use them as the benchmark dataset. 
Given a voice input, \textit{VoiceCoach} automatically recommends good voice modulation examples from the dataset
based on the similarity of both sentence structures and voice modulation skills.
Immediate and quantitative visual feedback is provided to guide further improvement.
\textcolor{black}{The expert interviews and the user study} provide support for the effectiveness and usability of \textit{VoiceCoach}.
\end{abstract}

\ccsdesc[500]{Human-centered computing~Human computer interaction (HCI)}
\ccsdesc[500]{Human-centered computing~Visualization}
\ccsdesc[500]{Human-centered computing~User interface design}

\keywords{\plainkeywords}

\printccsdesc

\input{introduction}

\input{relatedwork}
\input{requirements.tex}
\input{system.tex}

\input{expertinterview.tex}

\input{userstudy.tex}

\input{discussion.tex}
\input{conclusion.tex}

\input{acks.tex}

\balance{}

\bibliographystyle{SIGCHI-Reference-Format}
\bibliography{sample}

\end{document}

%% file: introduction.tex
\section{Introduction}
Public speaking is one of the most important interpersonal skills for both our everyday \textcolor{black}{lives} and careers. 
When delivering a public speech, voice is the primary channel for the speaker to communicate with the audience~\cite{yourspeakingvoice}. Therefore, voice modulation, the manipulation of vocal properties, has \textcolor{black}{a great influence} on audience engagement and the delivery of presentations~\cite{lucas2004art}.
Many studies~\cite{devito2003essential,lucas2004art,nikitina2011successful} have identified key elements for voice modulation including pitch, volume, pause, and speed. 
For example, increasing the speech speed can convey excitement, while slowing down and using appropriate pauses gives audiences \textcolor{black}{time to reflect on \textcolor{black}{the} speaker's words and form personal connections with the content.}
Higher volume 
leads to \textcolor{black}{a} vocal emphasis. A suitable repetition \textcolor{black}{in a} similar voice pitch \textcolor{black}{promotes clarity and enhances effectiveness in making key points}.
All these voice modulation skills have been proved critical to successful public speaking.

However, it is challenging to master and apply various voice modulation skills in public speaking.
Speakers can train themselves by following the general guidelines from the books \textcolor{black}{on} public speaking. However, this method suffers from the lack of immediate feedback, because it is often difficult for novice speakers to evaluate their \textcolor{black}{voice} accurately.
Another possible method is to join \textcolor{black}{a}
training programs and seek help from professional coaches.
However, the feedback from coaches could be subjected to their personal \textcolor{black}{preferences and be} inconsistent.
There lacks a quantitative method for evaluating speakers' performance and improvements in voice modulation skills.
Moreover, it remains unclear about how to combine different public speaking skills and adapt them to different speaking contents and presentation scenarios.

Several prior studies \cite{tanveer2015rhema,seetharaman2019voiceassist,rubin2015capture} have proposed computer-aided user interfaces to assist in voice modulation training by providing automated feedback on vocal properties such as volume. 
However, such feedback is determined by constant predefined thresholds, therefore failing to adapt to different speech contexts such as the content and the presentation purpose.
Besides, those systems do not provide concrete examples, which could make the learning process less effective.

To address the aforementioned challenges, we aim to develop an interactive system to support effective evidence-based training of voice modulation skills. We \textcolor{black}{work} closely with experienced public speaking coaches from an international training company for the past eight months to identify the challenges and detailed requirements. Our resulting system, {~\name}, helps speakers improve their voice modulation skills in four dimensions, i.e., \textit{pause}, \textit{volume}, \textit{pitch}, and \textit{speed}.
We first process 2,623 TED talk videos into over \textcolor{black}{300,000} audio segments based on sentences and build the benchmark of ``good'' examples for voice modulation. 
Then, we propose an effective recommendation approach to retrieve speaking examples for the speakers to explore and learn what can be improved in their voice. \textcolor{black}{The recommended examples are ranked by their similarity with the user input in terms of both speech content (i.e., text) and modulation patterns, 
facilitating the usage of the most appropriate voice modulation skills for the input sentence(s).}
This evidence-based training offers novice speakers \textbf{concrete and personalized guidelines} about what voice modulation skills can be used. 
Furthermore, during the practice of voice modulation skills, {~\name} provides \textbf{on-the-fly feedback} on the speaker's performance in terms of mastering the desired voice modulation skills, which is achieved by recognizing the differences between voice modulation skills in the speaker's speaking and the desired ones.
\textcolor{black}{Considering that} novice speakers may not necessarily have a background in visualization or even computer science, we propose \textbf{straightforward visual designs} and make them as intuitive as possible to convey the concrete guidelines and on-the-fly feedback for the training of voice modulation skills.

Our major contributions can be summarized as follows:

\begin{compactitem}

   
    \item We \textcolor{black}{present} an interactive visual system, {\name} ,
    to assist in the effective training of voice modulation skills in public speaking. {\name} can recommend concrete training examples from a TED talk database based on the semantic and \textcolor{black}{modulation} similarities, and supports on-the-fly quantitative feedback to guide the further improvement.
    

    \item We conduct 
    \textcolor{black}{expert interviews with professional coaches and a user study with university students,}
    which provide support for the effectiveness and usability of {\name} in facilitating the self-training of voice modulation skills.

\end{compactitem}

%% file: relatedwork.tex
\section{Related Work}
Our work is related to 
voice modulation, training systems for public speaking, and visualization of audio features.
\subsection{Voice Modulation}
Voice modulation refers to the manipulation of properties of voice~\cite{pisanski2016voice}, including pitch, volume, speed, etc. Researchers have conducted extensive studies on the voice modulation skills in the domain of public speaking, attempting to identify vocal techniques that contribute to a successful speech. Strangert~\cite{strangert2005prosody} analyzed speech behaviours of news announcers and politicians and summarized the characteristics of ``good'' speakers, where it was identified that pauses, changes of speed and dynamics of prosody made the speech efficient. Tsai \cite{tsai2015you} compared vocal characteristics of TED talkers with that of university professors and found out that TED speakers speak at a more consistent speed and with a deeper voice. Rosenberg et al.~\cite{hirschberg2005acoustic} examined the lexical and acoustic properties of charismatic speech. These studies shed lights on the effective vocal skills for public speaking. 
However, how to help users quickly and effectively train themselves to master these voice modulation skills still requires further exploration.





\subsection{Training Systems for Public Speaking}
Researchers in the HCI community have proposed several speech training systems, which offer automated feedback on users' speech quality. Many systems~\cite{damian2015augmenting, tanveer2015rhema, trinh2017robocop} evaluate speech quality by measuring vocal characteristics such as pitch, speech rate, and loudness. Their approaches quantify the quality by predefined thresholds regardless of sentences and contexts, which offers insufficient support for deliberate practice of particular sentences.
To address this problem, Narration Coach el al.~\cite{rubin2015capture} assisted users in recording a script by providing feedback on whether users satisfy voice modulation requirements that are specific to each sentence.
However, it requires users to specify those requirements such as spoken emphasis, which could be tedious and particularly difficult for novice users.
Therefore, our work studies how to automatically generate voice modulation strategies given an input script. Specifically, we analyze voice modulation strategies from \textcolor{black}{2,623} high-quality speeches (i.e., \textit{TED Talks}) which are used as the benchmark to recommend strategies.

Another key contributing factor of training outcomes is the feedback strategy. A large body of works has focused on providing in-situ feedback \cite{damian2015augmenting, dermody2016multimodal, schneider2015presentation, schneider2017presentation, tanveer2015rhema}. While such timely feedback is effective for immediate self-correction, long-term retention has been shown to be associated with intermittent feedback \cite{Schmidt89}.
Thus, another line of research proposes interactive systems for analyzing offline feedback to enhance self-reflection \cite{hoque2013mach, kurihara2007presentation, Tanaka15, zhao2017semi}. 
However, those systems only utilize simple charts with limited interaction support, and therefore are insufficient in helping users compare their performance and practice deliberately.  
Our work combines and extends both strategies by proposing a novel interactive visualization system to convey on-the-fly feedback, and by providing rich interactions to assist in analyzing performance in comparison with recommended examples in an iterative manner.

\subsection{Visualization of Audio Features}
Visualization is an intuitive and effective way of revealing patterns in audio. 
Much research has focused on developing visualization techniques to represent audio features. 
One of the most common methods is to use line charts to display temporal changes of feature values~\cite{mertens2004prosogram, zeng2019emoco}. Music flowgram \cite{jeong2016visualizing} extends line charts by introducing more visual elements such as color and height to encode features. 
Some works adopt matrix-like \cite{foote1999visualizing} or graph-based \cite{muelder2010content} visualization to describe the structural information of audio.
Others utilize metaphors such as clocks \cite{bergstrom2007conversation} and geographical maps \cite{pampalk2002content,morchen2005databionic}.

Considering the scenario of speech analysis, audio is often associated with words. Therefore, many visual systems have been developed to explore the relationship between audios and texts. The idea is to overlay audio features along with the scripts.
Prosograph \cite{oktem2017prosograph} horizontally aligns all the words with their corresponding prosodic features, enabling easy exploration of speech corpus.
VerseVis \cite{milton2015versevis} draws a filled-line graph, whose height encodes phonemes and color encodes accents. 
Patel and Furr \cite{patel2011readn} explores two ways of combining prosodic features with texts: one is to directly control properties of text, using horizontal position, horizontal placement and level of greyness to indicate duration, pitch and intensity respectively. The other is to augment text information by overlaying corresponding prosodic contours.

Although all these works ease the process of tracking temporal changes in the audio features, it requires extra time to both identify the repetitive patterns in lines of scripts and compare structural similarities of features. In comparison, our design gives a quick overview of frequent patterns in the audio collections by displaying technique combinations of varying lengths in a hierarchical order.
Furthermore, we convert continuous audio features into compact and intuitive glyphs to facilitate quick analysis of the similarity between lines of words.

%% file: requirements.tex
\section{Design Process and Requirement Analysis}
{\name} aims to help novice speakers understand, practice and improve their voice modulation skills. 
\textcolor{black}{To understand the current practice and challenges of the training process,
our design process started with an eight-hour training session offered by our industry collaborator, an international communication and leadership training company. During the training, we conducted contextual inquiries to collect information about the training process and difficulties encountered by trainees, which motivated the initial design of our system. In the later stages, we adopted an iterative development approach by carrying out bi-weekly meetings with four domain experts (\textit{E1-E4}) for eight months.
The experts are professional coaches from our industry collaborator, who all have at least six years' experience in the training of professional public speaking.
During the meetings, we collected experts' feedback on our early prototypes and updated the system design.
Similar to the system design process of prior research \cite{boyd2006participatory}, these experts serve as proxies to our target population in the requirement analysis and system design of {\name}. Specifically, the experts' expertise in public speaking helps us gain a deeper understanding of voice modulation skills.
Their experience of public speaking training makes them better aware of the difficulties that novice speakers may encounter in improving their modulation skills.
Also, they have deep insights into the limitations of traditional methods for training voice modulation skills.}

\textcolor{black}{The design requirements are formulated throughout the eight months and we summarize them as follows:}

{\textbf{R1.~Inform speakers of their voice modulation.}}
\textcolor{black}{All experts emphasized the importance of providing feedback to speakers on their performance of communication training, which is considered as the basis for improvements.
For example, \textit{E3} pointed out that the trainees usually overestimate the time they have paused when practicing the \textit{three-second pause} strategy, but underestimate their volume or pitch. 
Therefore, it is important to inform speakers 
of their usage of voice modulation skills.
}


{\textbf{R2.~Provide hints and evidence to guide potential improvements in speakers' voice.}} 
According to our expert interviews, another major challenge for novice speakers is how to practice and improve their voice modulation skills.
For instance,
\textcolor{black}{\textit{E4} said \textit{``Guidance is really important to novice speakers. They usually don't know how and when they need to use voice modulation skills.''}}
Thus, the system should help users quickly identify the issues or problems in their speech and further provide hints to guide their subsequent training based on their performance and preference. 


%

{\textbf{R3.~Illustrate the evidence with concrete examples.}} Our experts commented that they usually provide high-level tips such as \textit{``vary your tone more''}, \textit{``pause longer''} during the training session due to limited time.
Such tips, however, could be abstract and difficult for trainees to understand and apply correctly.
The system should provide concrete illustrations of voice modulation to promote efficient ``learning-by-examples''.

{\textbf{R4.~Enable on-the-fly feedback on speakers' vocal performance.}} 
During the iterative development process, we have found that users sometimes fail to make correct adjustments to their voice modulation when speaking the script, as it is often difficult to memorize all the details of their previous practices.
On-the-fly feedback could guide adjustments in a timely manner, making the practice more efficient and effective. 

{\textbf{R5.~Promote deliberate and iterative practice.}}
We have also observed that speakers could only focus on a few aspects during each practice. \textcolor{black}{\textit{E4} commented \textit{``Most people can't apply all types of modulation skills into one sentence and it is good enough to have two or three voice modulation skills on meaningful words or phrases.''} \textit{E1} said \textit{``We cannot expect people to master voice modulation at the first try.''} Therefore, the system should enable and encourage them to focus}
on specific types of voice modulation skills in an iterative manner, helping speakers practice and improve deliberately.





%% file: system.tex
\section{VoiceCoach}
\begin{figure}[!htb]
  \centering
  \includegraphics[width=1\columnwidth]{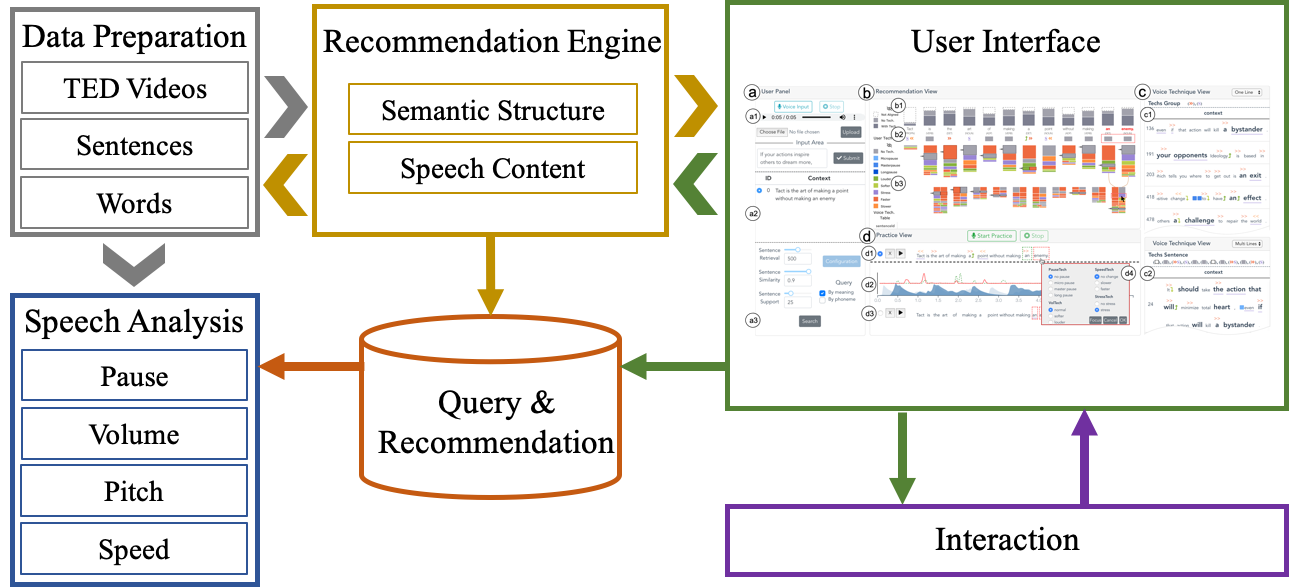}
  \vspace{-6mm}
  \caption{The system architecture of {\name}, which is comprised of four major modules, i.e., data preparation, speech analysis, recommendation engine, and user interface. 
}
  \label{fig:systemArchitecture}
\end{figure}

According to the aforementioned system requirements, we further design and implement {\name} \textcolor{black}{(Figure~\ref{fig:systemInterface})}, an interactive system for exploring and
\textcolor{black}{practicing voice modulation skills. The system architecture (Figure~\ref{fig:systemArchitecture}) consists of four major modules, i.e., data preparation, speech analysis, recommendation engine, and user interface.
The data preparation module creates the benchmark for voice modulation training. The speech analysis module analyzes modulation skills in users' audio input. The recommendation module retrieves good learning examples based on the input. The user interface module enables effective exploration and comparison of voice modulation skills in the retrieval results, and provides real-time quantitative feedback on users' performance. 
}



\subsection{Data Preparation}
The data preparation module aims to create a database of high-quality speeches that are used as the benchmark for training.
We choose TED Talks because they are widely considered as the pinnacle of public speaking \textcolor{black}{in terms of high-quality speech content and presentation skills \cite{gallo2014talk}. According to the official TED organizer guide \footnote{https://www.ted.com/participate/organize-a-local-tedx-event/tedx-organizer-guide}, the recording equipment is carefully set and tested to ensure a constantly good audio quality. The invited speakers have diverse professional backgrounds (e.g., entrepreneurs, educators) and the speeches cover over 400 topics.
Prior researches \cite{strangert2008makes,tsai2015you} have also used them as the benchmark for audio analysis of presentation styles.}

We collect videos from the TED Talk website\footnote{https://www.ted.com/talks} \textcolor{black}{published until June 2019.}
Then, audio clips of the videos are converted to scripts using the Amazon
Transcribe API\footnote{https://aws.amazon.com/transcribe/}.
A summary of the dataset is shown in Table~\ref{tb:data_statistic}.
\textcolor{black}{For each talk, the transcribed texts are split into sentences at periods, exclamation marks or question marks.
Each sentence contains all the spoken words together with their start and end time records.}

\begin{table}[!h]
\small
\caption{Dataset Properties.}
\vspace{-2mm}
\centering
\begin{tabular}{l l}
\hline
\textbf{Property}  & \textbf{Quantity}\\
\hline
Total number of talks & 2,623\\
Total length of all talks & 585.85 hours\\
Average duration of a talk & 13.4 minutes \\
Total word count & 5,350,391 \\
Total sentence count & 334,692 \\
Average words of a sentence & 15.99\\
Total topic categories & 430 \\
Total speakers' occupations & 447 \\
\hline
\end{tabular}
\label{tb:data_statistic}
\end{table}


\subsection{Speech Analysis}
\textcolor{black}{After a user uploads his/her audio input, the speech analysis module will process the audio and detect the employed modulation skills in terms of four vocal properties (i.e., pause, volume, pitch, and speed):}

\textbf{Pause}: We focus on intentional pause other than unnecessary interruptions. We calculate it by measuring the interval between two words, which are classified according to coaches' training specifications  - [0.5s, 1s): ``brief pause'', [1s, 2.5s): ``master pause'', and [2.5s, $\infty$): ``long pause''.

\textbf{Volume:} We compute the average volume for each word, as well as the average and the standard deviation (SD) for each sentence. 
Then, we 
label words that are louder ($>$ 1.1 times or $>$ 1 SD) 
than the sentence as ``louder'' and softer ($<0.67$ times or $< -1$ SD) than the sentence
as ``softer''.

\textbf{Pitch:} A higher pitch relates to a vocal stress. Similar to volume calculation, we track the pitch contours to find peak values. Specifically, we label words that are higher pitched ($>$ 1.25 times) or have more pitch variation ($>$ 1 SD) as ``stress''.

\textbf{Speed:} We consider two variations of speed 
(i.e., faster and slower).
We compute the Syllables Per Minute (SPM) for each word, as well as the average and standard deviation of SPM for each sentence.
Then, we label those that are faster ($>$ 1.5 times) or have more variation ($>$ 1 SD) as 
``faster''
and slower ($<$ 0.67 times) or have more variation ($<$ -1 SD) as 
``slower''.

We set
the above default thresholds empirically together with our experts, and also allow users to change \textcolor{black}{them} in the user panel (Figure~\ref{fig:systemInterface}(a3)) to enable interactive customization by users when necessary.

\subsection{Recommendation Engine}
\textcolor{black}{The recommendation module retrieves TED speech examples from the TED dataset by considering both the semantic structure of speech content and the voice modulation skills employed in the user input. It consists of three phases.} The first phase is to search semantically relevant sentences in the database. We leverage the state-of-the-art sentence encoder method \cite{cer2018universal} to embed sentences into feature vectors that preserve the semantic information. Then,~\textcolor{black}{the recommendation module finds examples that are close to the query based on cosine similarities in the high-dimensional embedding space. 
To speed up the search, we leverage Annoy \footnote{https://github.com/spotify/annoy}, which is one the most popularly-used nearest neighbor search libraries and has been used in the recommendation engine in Spotify\footnote{https://www.spotify.com/}.}
\textcolor{black}{Hence, we retrieve a set of semantically relevant sentences from the dataset.} The second phase is to align the sentence of user input with the retrieved sentences based on structural information. The retrieval results from the first phase~\textcolor{black}{will be aligned} with the input query based on part-of-speech features to facilitate the comparison of sentence structures and voice modulation skills employed in the corresponding sentences.
The third phase is to search frequent modulation combinations based on aligned words in the retrieved sentences. At this phase, \textcolor{black}{the recommendation module} recommends the usage of voice modulation based on n-grams, which incorporates different lengths of word contexts. It constructs a FP tree~\cite{han2000mining} on the \textcolor{black}{structurally aligned technique sequences of} retrieved examples, and finds frequent \textcolor{black}{voice modulation} combinations in the tree. A high support threshold value decreases the generated combinations, while a low one reserves more unusual combinations. The default value is 0.05, which can be interactively adjusted by users in the user panel (Figure~\ref{fig:systemInterface}(a3)).

\subsection{User Interface}

\textcolor{black}{To make the voice modulation training more user-friendly, we design an interactive visual analytics system called {\name} \textcolor{black}{(Figure~\ref{fig:systemInterface})} with four coordinated views, including (a) {\up}, (b) {\rv}, (c) {\vtv}, and (d) {\pv}.
The voice modulations are 
visually encoded by both colors and glyphs, as shown in Table~\ref{tb:ted_statistic}.}



\begin{table}[!h]
\small
\caption{Glyph Encoding of Voice Modulations}
\vspace{-2mm}
\centering
\begin{tabular}{lll}
\hline
Property                & Modulation   & Glyph  \\ \hline
\multirow{2}{*}{Speed}  & Faster     & {\textcolor{fast}{\faAngleDoubleRight}} \\ \cline{2-3} 
                        & Slower    & {\textcolor{slow}{\faAngleDoubleLeft}} \\ \hline
\multirow{2}{*}{Volume} & Louder       & {\textcolor{louder}{\faLevelUp}} \\ \cline{2-3} 
                        & Softer       & {\textcolor{softer}{\faLevelDown}}     \\ \hline
\multirow{3}{*}{Pause}  & Brief pause  & {\textcolor{micropause}{\faSquare}} \\ \cline{2-3} 
                        & Master pause & {\textcolor{masterpause}{\faSquare\ \faSquare}} \\ \cline{2-3} 
                        & Long Pause   & {\textcolor{longpause}{\faSquare\ \faSquare\ \faSquare}} \\ \hline
Pitch                   & Stress        & {\textcolor{stress}{\underline{\textbf{S}}}} \\ \hline
None                    & No Tech.       & {\textcolor{normal}{\faSquare}} \\ \hline
\end{tabular}
\label{tb:ted_statistic}
\end{table}

\subsubsection{User Panel}
The {\up} accepts different types of user input including audio streaming, audio files, and texts in Figure~\ref{fig:systemInterface}(a1). Then, it presents the resulting sentences in a table in Figure~\ref{fig:systemInterface}(a2). The user can adjust the parameters of both the example retrieval algorithms and the voice modulation detection approach at the bottom (Figure~\ref{fig:systemInterface}(a3)).



\subsubsection{Recommendation View}
\textcolor{black}{After the system analyzes the user input and retrieves ``good'' examples from the benchmark dataset, the {\rv} (Figure~\ref{fig:systemInterface}(b1)(b3)) is designed to summarize different combinations of voice modulation skills and provide speakers with hints for further improvements.
}

\textcolor{black}{We propose a stacked bar chart based design to visualize information of voice modulation skills in a coarse-to-fine manner (Figure~\ref{fig:systemInterface}(b1)(b3)).
}
\textcolor{black}{The top part (Figure~\ref{fig:systemInterface}(b1)) presents general summary of retrieval results with respect to three conditions (i.e., not aligned, no technique and with technique). Each condition is encoded by a color or texture. The height of each segment of the stacked bar indicates the frequency of each type of the three conditions. For example, 
a tall dark gray bar implies that its corresponding word is popular for modulation, while a tall dashed-outlined bar indicates that the recommendation results of the word have insufficient support.}
\textcolor{black}{
The n-gram-based hierarchical visualization (Figure~\ref{fig:systemInterface}(b3)) summarizes the varied-length combinations of voice modulation skills
in the retrieval results.
The first row of stacked bars in Figure~\ref{fig:systemInterface}(b3) visualizes the voice modulation skills of each word, where a stacked bar chart is displayed under each word.
The second row of the stacked bars in Figure~\ref{fig:systemInterface}(b3) shows the frequent combination of voice modulation skills for two adjacent words.
The stacked bars are horizontally aligned at the center of the corresponding two words. 
The bars within each stacked bar are sorted in a descending order by the frequency of the corresponding voice modulation skill or the combination of voice modulation skills.
When the user hovers over a voice modulation combination, its corresponding words will be highlighted in bold red.
}

\begin{figure}[!htb]
  \centering
  \includegraphics[width=1\columnwidth]{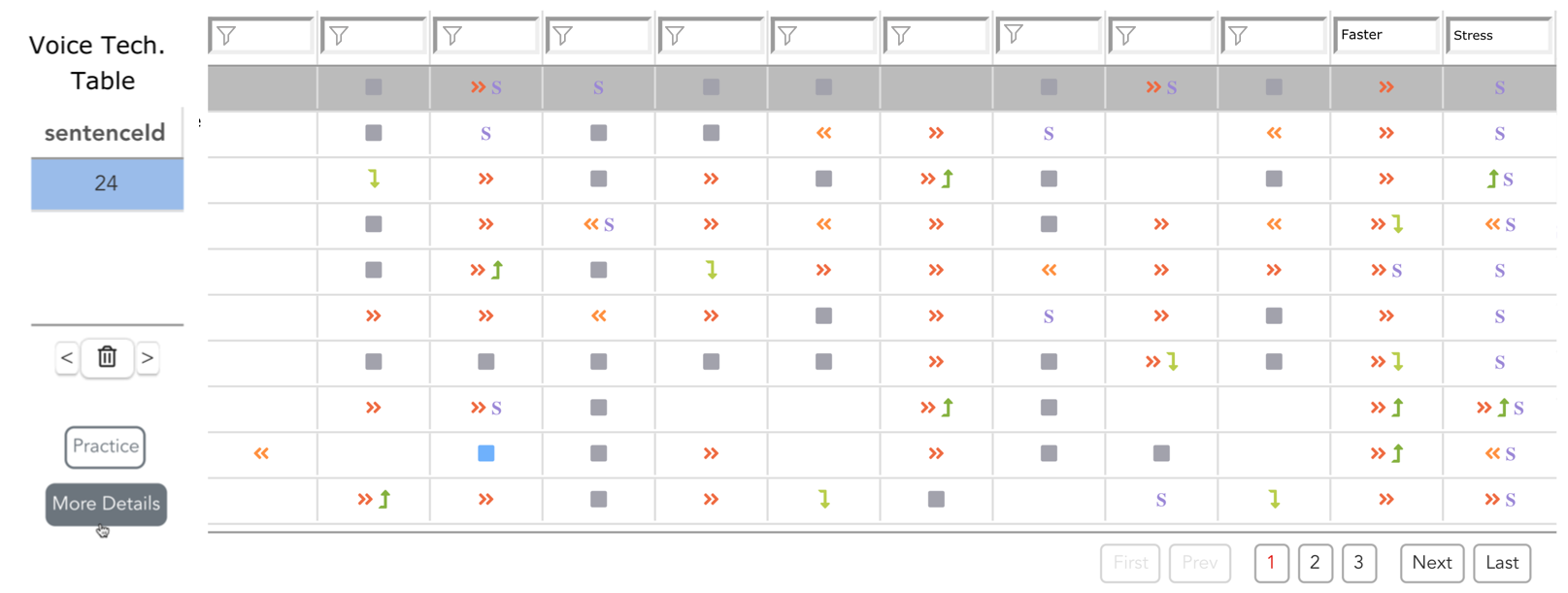}
  \vspace{-7mm}
  \caption{The voice technique table.
  Users can filter sentences with specific voice modulations (e.g., faster and stress), then select some corresponding sentences for further exploration in the voice technique view by clicking the detail button.}
  \label{fig:voiceTechTable}
\end{figure}

\textcolor{black}{
The \textit{voice technique table} at the bottom (Figure~\ref{fig:voiceTechTable}) shows a list of voice modulation skill sequences employed by the speaking examples in the TED benchmark dataset. These voice modulation skill sequences are sorted by their 
similarity 
with the voice modulation skill sequence extracted from the speaker's voice input. We use Hamming distance to measure the similarity between two sequences.
The speaker can further explore interesting combinations of voice modulation skills by filtering techniques at the header of the table.
}

To ease comparison between the voice modulation skills employed by a speaker and the modulation skills used in the TED benchmark dataset, we come up with three designs to enhance the {\rv}. First, the modulation skills of a speaker (Figure~\ref{fig:systemInterface}(b2)), which are set as the baseline for comparison, are encoded by colored glyphs. Second, arrow markers are added in the n-gram based visualization to highlight the modulation skills used by both a speaker and the TED talks. Third, some buttons ({\faEye and {\faEyeSlash}}) in Figure~\ref{fig:systemInterface}(b3) are added to help a speaker interactively set whether the n-gram-based visualization is shown or not. When it is hidden, the voice technique table will automatically move up and be positioned close to the glyphs of the voice modulation sequence of the speaker, enabling sequential comparative analysis of the voice modulation patterns.

\subsubsection{Voice Technique View}
\textcolor{black}{
When a user clicks on the voice modulation of his/her interest in the {\rv}, {\name} can retrieve the TED talk segments that use the desired voice modulation skills and list them in the {\vtv}.
These TED talk segment examples are ranked by the sentence similarity of both sentence structural and semantic meanings between the user input and the TED talk segment examples.
The retrieved TED talk segment examples can be highlighted in one-line mode~(Figure~\ref{fig:systemInterface}(c1)) or multi-line mode~(Figure~\ref{fig:systemInterface}(c2)), which enables the user to quickly locate and compare the local context of different voice modulation skills.
When the user clicks on a word or a sentence ID in the {\vtv}, the corresponding original TED talk voice will be played to give users a concrete understanding of the voice modulation skills.}

\subsubsection{Practice View}
\textcolor{black}{The {\pv} consists of three components:
(1) a reference example showing the sentence with highlighted techniques that the user wants to practice (Figure~\ref{fig:systemInterface}(d1)), 
(2) a real-time feedback chart providing immediate quantitative feedback on the voice modulation skills employed by the speaker in his/her practice (Figure~\ref{fig:systemInterface}(d2)), and (3) a practice collection (Figure~\ref{fig:systemInterface}(d3)) storing and displaying all recorded practices 
To promote deliberate practice, the speaker is allowed to customize the words to focus on and techniques to be improved (Figure~\ref{fig:systemInterface}(d4)).
To provide real-time and quantitative feedback on voice modulation, the feedback chart updates the real-time value of pitch (red solid line) and volume (dark blue area) of the current practice simultaneously, while the volume (light blue area) and pitch (green dashed line) of the previous practice are set as the baseline. Other vocal properties can also be inferred from the chart. For example, segments with zero volume indicate the pause and the speed of the volume wave suggests the speech rate.}

\subsection{Usage Scenario}
We describe how Andy, an undergraduate student, utilizes {\name} to practice and improve his voice modulation skills.
Andy is preparing for a speech about negotiation skills, and he decides to take Isaac Newton's famous quote - ``\textit{Tact is the art of making a point without making an enemy}'' - as a highlight of his talk.
Therefore, he refers to {\name} to perform deliberate practices on this quote.  

After recording the script, he examines the {\rv} which shows the voice modulation skills he applied, in comparison with the recommended results. 
As shown in Figure~\ref{fig:systemInterface}(b2), he quickly notices that the voice modulation skills he used for several words (i.e., the color rectangles indicated by a black arrow on their left) are consistent with those recommendation results (e.g., \textit{``tact'', ``art'', ``of'' and ``point''}), but there are also words where he does not use any voice modulation skills (indicated by the gray rectangles with a black black arrow) while TED speakers employed certain voice modulation skills. More specifically, an obvious exception of his speaking lays in the phrase ``\textit{making an enemy}'', which is a key part of this quote but no voice modulation skills are adopted by Andy.
He first tries to improve his speaking for ``\textit{an enemy}'' by applying some voice modulation skills to them.
Since the most frequent combination (i.e., {\textcolor{fast}{\faAngleDoubleRight}} ``faster'' and {\textcolor{normal}{\faSquare}} ``no tech'') does not apply any technique to the word \textit{``enemy''}, he chooses the second most popular voice modulation combination, i.e.,  {\textcolor{fast}{\faAngleDoubleRight}} ``faster'' and {\textcolor{stress}{\faSquare}} ``stress''.


He decides to find an example with those techniques to mimic the voice modulation.
He applies filters in the \textit{voice technique table} to query sentences from the database. 
After he clicks the first returned result that has the highest similarity score with his phrase input, and listen to the example to \textcolor{black}{develop} a concrete idea about how a voice modulation combination of  {\textcolor{fast}{\faAngleDoubleRight}} ``faster'' and {\textcolor{stress}{\faSquare}} ``stress'' should be, as shown in Figure~\ref{fig:systemInterface}(c).

\begin{figure}[!htb]

  \centering
  \includegraphics[width=1\columnwidth]{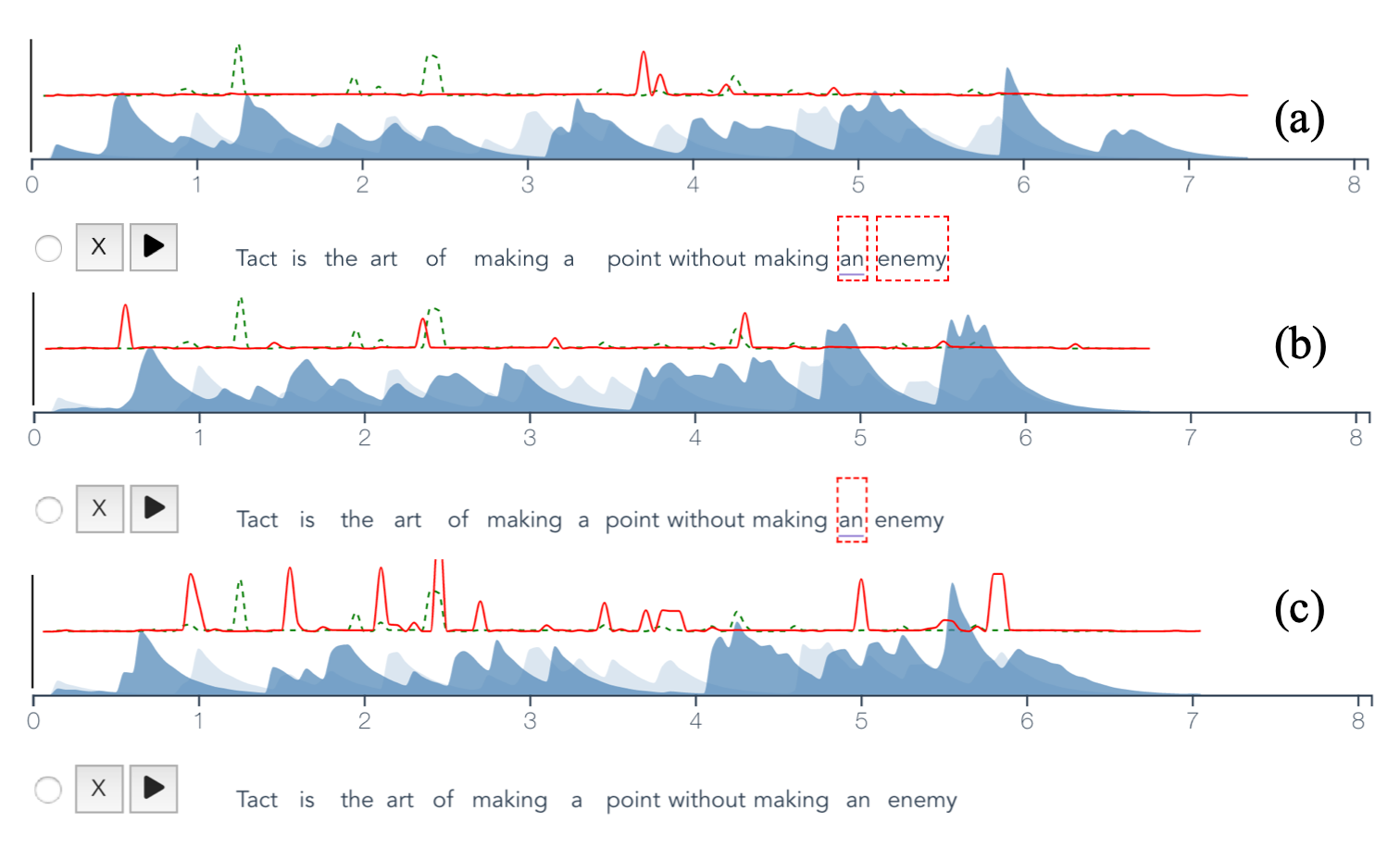}
  \vspace{-7mm}
  \caption{An illustration of multiple practices by a user, where the user focuses on practicing the phrase ``an enemy''.
  The decreasing number of the dashed red rectangles from (a) to (c) show the user's improvement of the voice modulation skills.
  }
  \label{fig:practiceProcess}
\end{figure}

Andy further uses the practice view to improve his speaking. As shown in Figure~\ref{fig:practiceProcess}(a-c), his volume (the dark blue area) and pitch (the red line) are detected and shown in real time in each of his speaking practice for this quote. The corresponding volume (the light blue area) and pitch (the dotted green line) of his original speaking are used as a reference to show his improvement in each practice. The inconsistency of voice modulation for the phrase  ``\textit{an enemy}'' between each practice and the selected voice modulation combination is also highlighted in red dashed rectangles on the original text. From Figure~\ref{fig:practiceProcess}(a-c), it is clear to see that Andy correctly applied the voice modulation combination of  {\textcolor{fast}{\faAngleDoubleRight}} ``faster'' and {\textcolor{stress}{\faSquare}} ``stress'' into the phrase  ``\textit{an enemy}'' after three rounds of practice.

%% file: expertinterview.tex
\section{Expert Interview}
\textcolor{black}{We performed in-depth interviews with three domain experts (i.e., \textit{E1-E3}), who also participated in our requirement analysis interviews, to evaluate the effectiveness and usability of {\name}.}
We started the interviews by explaining the functions and visual encodings
of {\name}. \textcolor{black}{A usage scenario was also introduced to showcase the usage of {\name}.}
Then, we asked the experts to freely explore our system in a think-aloud manner and finish their exploration tasks, 
\textcolor{black}{e.g., examine the recommendation results according to their voice input, select one desired modulation for further practice, and iteratively
practice with on-the-fly quantitative feedback.}
After that, we collected their feedback on {\name}.  
Each interview took about 1 hour,
and all the interviews were recorded with the experts' consent.
Overall, the experts showed great interest in {\name}. Their feedback was summarized as follows.

\textbf{Usefulness} \textcolor{black}{All the experts agreed that \textcolor{black}{the evidence-based training in {\name}} could be 
helpful for novice speakers to improve their voice modulation skills.}
\textcolor{black}{\textit{E1} and \textit{E3} mentioned that novice speakers are often not sure about what voice modulation skills to use and how to combine them in a new script,}
\textcolor{black}{even though they may also already be aware of some high-level tips for voice modulation skills.}
\textcolor{black}{They thought our recommendation strategy was new and clever. The voice modulation examples recommended by {\name} provide speakers with evidence-based guidance. They can select suitable modulation skills for different sentences.}
\textcolor{black}{\textit{E2} pointed out that the on-the-fly feedback provided by {\name} was more useful than that of traditional training, as there is usually just one coach with multiple students in a class of a traditional training program, making it difficult for the coach to provide sufficient and timely feedback to every student.
\textit{E1} commented that the quantitative feedback in {\name} is very helpful for enabling a user to master voice modulation skills, as it provides the user with concrete real-time evaluations of their voice modulation skills during their practices.
}
During the interviews, 
\textcolor{black}{one interesting finding was that different coaches could have very different preferences for voice modulation. For instance, \textit{E1} mentioned that pause is one of the most important and
difficult voice modulation skills, thus his training often focused on pauses.
On the contrary, \textit{E2} confidently emphasized that the art of a successful speech 
lay in the good modulation of the volume and speed.} Such observations further confirmed the importance of the example recommendations in {\name}, which provide students with the flexibility to choose and follow the suitable ``good'' voice modulation examples.


\textbf{Visual designs and usability}
\textcolor{black}{All three experts appreciated the evidence-based training provided by {\name}. 
They confirmed that the overall visualization designs were intuitive and easy to understand.
For the {\rv}, \textit{E1} said that it demonstrated the diversity of voice modulation skills. \textit{E2} pointed out \textit{``Though the {\rv} seems to be the most complex view of {\name} at first glance, I can quickly understand and learn how to use it after your brief introduction.}'' All experts mentioned that most of the top-ranked recommendation examples in the {\vtv} made sense to them. 
By clicking the corresponding sentence in the {\vtv}, they could conveniently check how those voice modulation skills were used by the TED speakers. For the {\pv}, they agreed that the real-time feedback charts, as well as the highlighted text boxes, help them recognize the difference between different practices.
In addition, the experts were highly impressed by the convenient and smooth interactions of {\name}.
}

\textbf{Limitations and suggestions} Despite the overall positive feedback from the experts, they also pointed out some limitations of {\name} and gave us insightful suggestions on it.
\textit{E2} said that {\name} currently \textcolor{black}{only} recommended ``good'' voice modulation examples for \textcolor{black}{speakers} to follow, \textcolor{black}{while speakers} could also benefit from negative examples. \textcolor{black}{By informing \textcolor{black}{them} of ``bad'' modulation such as a monotone voice, they could easily know what mistakes they should avoid.}
\textit{E2} suggested that it would be interesting to 
\textcolor{black}{classify}
the speakers into different types (e.g., fast speaker vs. slow speaker, soft speaker vs. loud speaker) 
and to deliberately recommend voice modulation examples to them (e.g., recommend fast speaking examples to slow speakers and loud speaking examples to soft speakers). 
Due to the limited voice modulation datasets that are available, we have left this as part of our future work.




%% file: userstudy.tex
\section{User Study}

We conducted a well-structured user study to evaluate the effectiveness and usability of {\name} for the training of voice modulation skills. 
Since the concrete voice modulation examples (the {\rv} and {\vtv}) and the immediate and concrete feedback (the practice view) are the two major desirable functions of {\name}, we designed the user study with an emphasis on
these two aspects. 
Specifically, we aimed to answer the following questions: 
\begin{itemize}
    \setlength{\itemsep}{0pt}
    \setlength{\parsep}{0pt}
    \setlength{\parskip}{0pt}
    \item \textbf{Recommendation helpfulness:} How helpful is our system in finding appropriate voice modulation examples to guide the practice?
    \item \textbf{Effectiveness of immediate feedback:} How effective is our system for improving participants' voice modulation skills in terms of getting quick and quantitative feedback?
    \item \textbf{Overall usability \textcolor{black}{and effectiveness}:} \textcolor{black}{Is {\name} effective for improving participants' skills of voice modulation and is it easy to use?}
\end{itemize}



\subsubsection{Participants}

We recruited 18 \textcolor{black}{university} students (4 females, $age_{Mean} = 23$, $age_{SD} = 2.52$) from a local university through word-of-mouth and flyers.
They came from different backgrounds, including chemistry, math, computer science, mechanical engineering, and finance. Each participant received \$17. All the participants had experiences of public speaking, but none of them had attended any professional training of voice modulation. They have all expressed the desire to improve their presentations and an eagerness to improve their skills of voice modulation. All the participants had normal vision and hearing.

\subsubsection{Experiment Design}

Before the study, we worked together with the coaches (\textit{E1}, \textit{E2}) and selected 13 sentences (\textit{S1-S13}) as training examples.
These examples have been popularly used in their training programs.
\textcolor{black}{Our user study consisted of four sessions.
}

In the first session,
we introduced the purpose and the procedures of our study. After that, we illustrated the skills of voice modulation that we have mentioned in this paper with example videos and gave them some general tips about the usage of such skills. 
After they had grasped the concepts of voice modulation, we demonstrated how to use our system.

\textcolor{black}{In session two, we asked participants to freely explore {\name} in a think-aloud manner with four sample sentences \textit{(S1-S4)}, which aimed to familiarize} them with the system.

In session three, participants were presented with another \textcolor{black}{five} sentences \textit{(S5-S9)} and asked to examine the results generated by the {\rv} and the {\vtv}.
The tasks were to explore the recommended voice modulation skills and their corresponding words or phrases.
Meanwhile, they were requested to report how many recommended examples they believed were relevant in terms of sentence structure and voice modulation skills among the top five retrieval results in the {\vtv}. 
Their click activities were also captured for further analysis. 
At the end of session three, participants needed to complete a questionnaire consisting of 11 questions, where they evaluated the recommendation results (\textit{Q1-11}) in a 7-point Likert scale, as shown in Table \ref{table:questionnairetable}.

\textcolor{black}{In Session four}, we compared our system with a baseline system using another set of four sentences \textit{(S10-S13)}, where the baseline system was a simplified version of {\name} by removing the feedback generated from the {\pv} and only reserved the functions of recording and playback. This simplified system only allowed \textcolor{black}{participants} to listen to his/her own audios and make the adjustment accordingly, which simulated \textcolor{black}{real-world practice.}
For each sentence, \textcolor{black}{participants} were asked to practice it with the same pre-defined instructions using either {~\name} or the baseline system. To minimize the learning effect,
we evaluated the two systems in a counterbalanced order. 
Also, a questionnaire of 5 questions (\textit{Q12-Q16}) in Table \ref{table:questionnairetable} with a 7-point Likert scale was to be finished afterwards. 
After four sessions, we conducted a post-study survey with the participants, during which we had them finish (\textit{Q17-\textcolor{black}{Q25}}) in Table {\ref{table:questionnairetable}} and answer some general questions about their experience of the training. 
\textcolor{black}{The whole study lasted about 90 minutes.}

\begin{table}[!htb]
 \caption{Three questionnaires designed for Sessions three, four and the post-study survey. Assessment of the quality of of the recommendation results in four aspects: informativeness (\textit{Q1-Q3}), visual design (\textit{Q4-Q6}), decision making (\textit{Q7-Q9}), usability (\textit{Q10-Q11}). Assessment of the effectiveness of vocal practice: self-awareness (\textit{Q12-Q13}), self-adjustment (\textit{Q14-Q15}), self-reported evaluation (\textit{Q16}). Participants' feedback about {\name}: voice modulation (\textit{Q17-Q18}), system components (\textit{Q19-Q22}), usability (\textit{Q23-Q25}).}  
\centering
    \vspace{-2mm}
  \includegraphics[width=1\columnwidth]{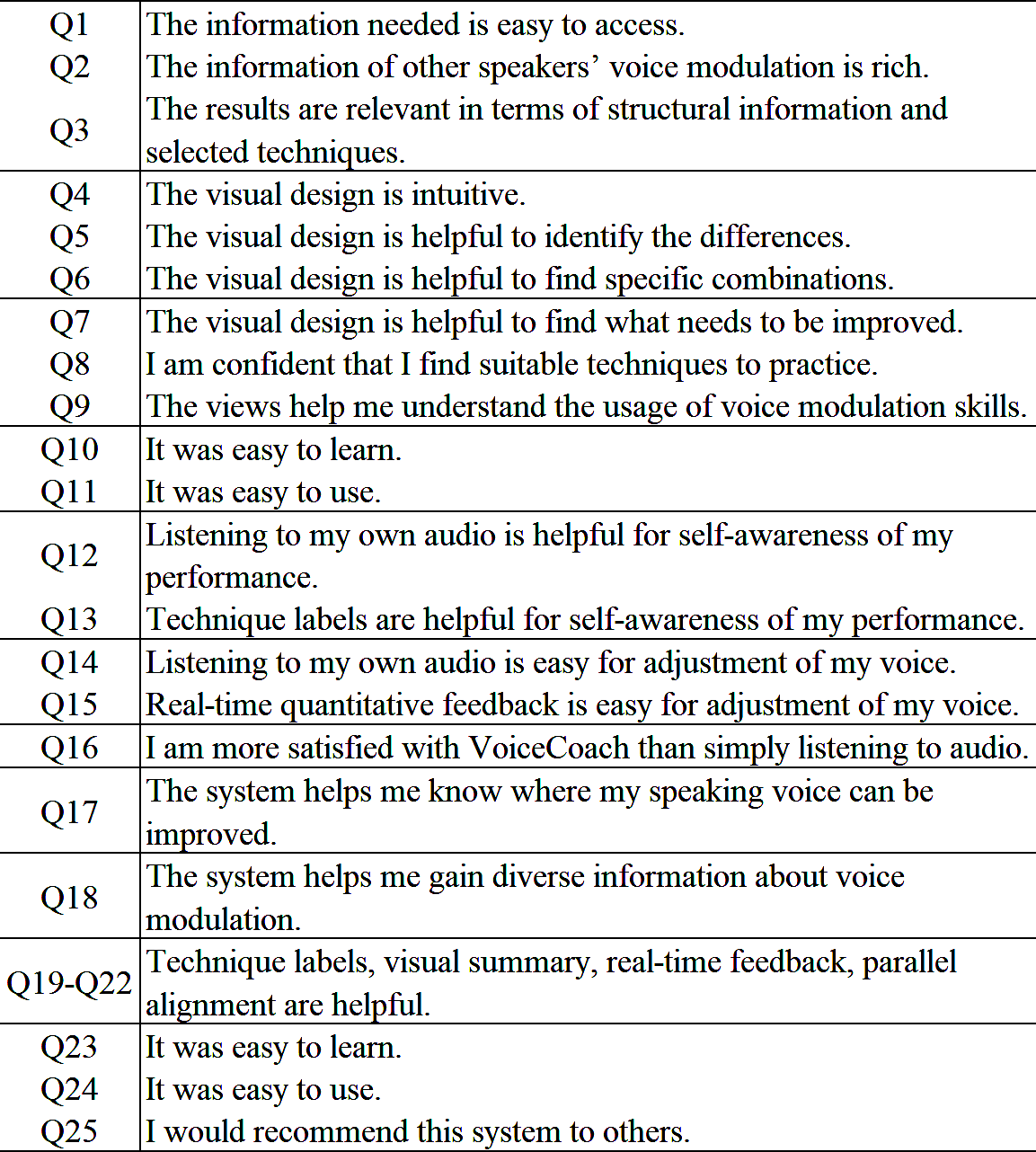}
  \label{table:questionnairetable}
  \vspace{-6mm}
\end{table}

\subsection{Results and Analysis}




\subsubsection{Evaluation on Recommendation Results}
\begin{figure}[!htb]
  \centering
  \includegraphics[width=1\columnwidth]{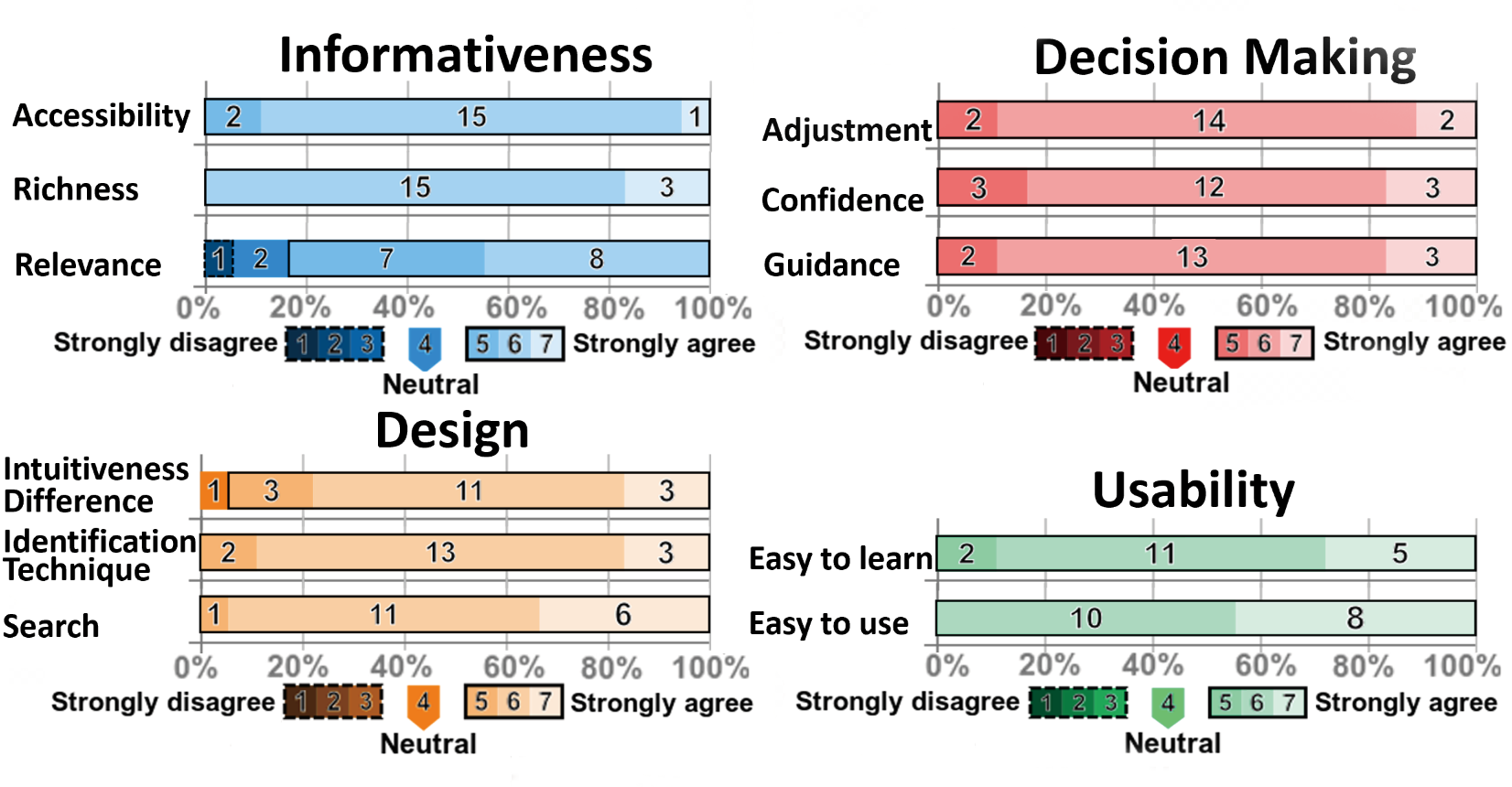}
  \vspace{-8mm}
  \caption{The results of questionnaire about helpfulness of recommendation in four aspects, including informativeness (\textit{Q1-Q3}), decision making (\textit{Q7-Q9}), design (\textit{Q4-Q6}), and usability (\textit{Q10-Q11}).}
  \label{fig:recommendation}
\end{figure}

\textcolor{black}{We analyzed the user-generated data (i.e., click data, report of the number of relevant examples in the top 5 retrieved results) and the ratings from the questionnaire (\textit{Q1-Q11} in Table{\ref{table:questionnairetable}}).
The results show
the usability and effectiveness of \textcolor{black}{the {\rv} and the {\vtv}}.}
On average,
participants 
clicked on modulation combinations \textcolor{black}{in the {\rv}} about 4.27 times ($SD = 1.27$) before they 
\textcolor{black}{settled down to a desirable combination,}
and \textcolor{black}{4.21 ($SD = 1.21$)} of top 5 retrieved results \textcolor{black}{displayed in the {\vtv}} satisfied the participants' needs. 
The relevance rate of the recommended examples was 89\%.
Besides, most participants showed positive responses to the recommendation results, especially in terms of decision making and usability. 
Interestingly, one participant (5.6\%) disagreed about the relevance of the recommendation results and one participant (5.6\%) was neutral about the intuitiveness of the visual design of {\rv}. 
The summary of the feedback is shown in Figure \ref{fig:recommendation}.
In addition, we had our experts \textit{E1, E2} go through all the chosen techniques of participants, and they found them reasonable and applicable.

\subsubsection{Evaluation on Practice View}
\begin{figure}[!htb]
  \centering
  \includegraphics[width=1\columnwidth]{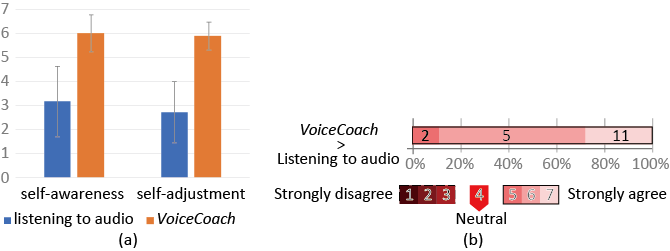}
  \vspace{-6mm}
  \caption{The results of questionnaire about user experience of practice. (a) Comparison of self-awareness and self-adjustment between {\name} and the baseline. (b) Participants' responses to \textit{Q16}.}
  \label{fig:practiceview}
  \vspace{-3mm}
\end{figure}

\textcolor{black}{We ran Wilcoxon signed-rank tests on the feedback for self-awareness and self-adjustment during the voice modulation practice in Session four,
to compare the effectiveness between {\name} and the baseline.}
The result (Figure \ref{fig:practiceview} (a)) shows a significant difference in the self-awareness scores ($p < 0.001 $, $Z = -3.75$), which indicated that {\name} better helped participants understand their vocal performance ($Mean = 6.00$, $SD = 0.77$) compared with listening to personal audio records ($Mean = 3.17$, $SD = 1.47$). Significant differences in the self-adjustment scores were also observed ($ p < 0.001$, $ Z = -3.70$), showing that {\name} better assisted participants in adjusting the participants' voice ($Mean = 5.89$, $SD = 0.58$) compared with the baseline method ($Mean = 2.72$, $SD = 1.27$).

Furthermore, \textcolor{black}{the ratings of} the questionnaire (Figure \ref{fig:practiceview} (b)) suggest that all participants prefer {\name} for practicing.

\begin{figure}[!htb]
  \centering
  \includegraphics[width=1\columnwidth]{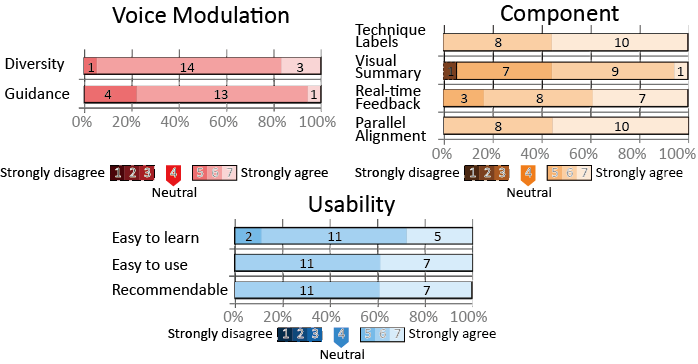}
  \vspace{-7mm}
  \caption{The results of participants' feedback about system in terms of voice modulation (\textit{Q17-Q18}), component design (\textit{Q19-Q22}) and usability \textit{(Q23-Q25)}.}
  \label{fig:systemuse}
  \vspace{-3mm}
\end{figure}

\subsubsection{Overall feedback on {\name}}
\textcolor{black}{We collected feedback from the questionnaire (\textit{Q17-Q25} in Table \ref{table:questionnairetable}) in the post-study survey.
The result (Figure \ref{fig:systemuse}) shows that}
all participants found that {\name} enriched their knowledge about voice modulation and helped them improve vocal skills. Also, most participants agreed that the four core components of the system were useful, especially technique labels and parallel alignment. One participant (\textit{P2}, Female, 24) found the visual summary less helpful, because it took her a while to understand the design of {\rv}. In general, participants claimed that {\name} had good usability.


During the post-study interview, we asked participants about their experience of using {\name} and the new knowledge of voice modulation they had learnt. We also collected comments and suggestions for
the {\up}, the {\rv}, the {\pv} and the {\vtv}.

\textbf{Training experience} In general, all participants felt excited about {\name}. One participant (\textit{P8}, Male, 20) strongly believed \textit{``({\name}) would be a helpful training tool for the speakers to practice their voice anytime anywhere.''} Five participants mentioned that by observing the differences between their voice and experts' voice, they gained insights about their inadequate usage of voice modulation. One of them (\textit{P4}, Male, 23) commented \textit{``...comparing with TED talkers made me realize that I normally spoke very fast and did not vary the speed much."} Another participant (\textit{P13}, Female, 18) was amazed by the power of a pause after her training: \textit{``I can't believe that a pause has such magic to make my voice sound so dramatic.''}

\textbf{Visual designs}
The visual design of the system seemed intuitive to most participants, especially the technique labels for feedback. One (\textit{P18}, Male, 22) said \textit{``The technique labels were simple and compact. Instead of listening to audio myself, I could quickly discover the sequential patterns (of voice modulation) in audio by these labels.''} Many participants found the arrow markers in the {\rv} helpful for identifying the differences between their voice and others' in all levels of voice modulation combinations. Interestingly, we noticed that some of them held contradictory opinions towards the sentence-level summary of voice techniques. One participant (\textit{P2}, Female, 24) found it less useful than n-gram-based visualization: \textit{`` The sentence I selected in the voice technique table did not seem relevant to my sentence.''} While another (\textit{P13}, Female, 18) thought \textit{`` The examples recommended in the table were so helpful for learning.''} The conflicts may be caused by the limited size of our dataset. 
One participant (\textit{P13}, Female, 18) described the practice as a voice game: \textit{``It was very interesting to see the real-time feedback of my voice on the screen. It reminded me of where and when I should make adjustments.''} Also, she expressed her difficulty in focusing on several dimensions simultaneously during the practice.

\textbf{Interactions} 
Overall, participants enjoyed the rich and effective interactions which helped them explore recommendation results. Many participants mentioned about the convenience of parallel alignment and the auto-focus of the corresponding contexts of selected techniques in the voice technique view. One (\textit{P14}, Male, 28) commented \textit{``The voice technique view saved my time. I could discover combinations of interests by one glance at the table.''} Another one (\textit{P4}, Male, 23) added \textit{``It was very considerate of you to let me listen to the words I want with a simple click. I didn't bother to listen to the whole sentence.''} Many participants agreed that it was beneficial to let them focus on specific words and modify the unwanted techniques, which eased the whole process of practices.
After the experiment, two participants showed their strong interests in {\name} and spent extra time on exploring our system.



%
\subsubsection{\textcolor{black}{Evaluation by coaches}}
\textcolor{black}{To further determine the training effectiveness of {\name}, we invited the two aforementioned coaches (\textit{E1, E2}) to evaluate the speakers' performance. Specifically, we recruited another 24 \textcolor{black}{university} students (9 females, $age_{Mean} = 24$, $age_{SD} = 2.47$) from our university and randomly divided them into two groups (\textit{G1}, \textit{G2}). Participants were asked to practice their voice modulation skills based on the same script with or without {\name}. The script was a 30-second speech opening pre-selected by \textit{E1} and \textit{E2}, and \textit{G1} was set as the control group. After that, coaches evaluated speakers' final audio presentation in terms of diversity, coherence, and expressiveness of voice modulation with a 7-point Likert scale. Both coaches were blind to the study condition.}

\textcolor{black}{We 
analyzed the performance scores of \textit{G1} and \textit{G2} using Wilcoxon signed-rank tests. There was a significant difference ($ p = 0.03 $, $ Z = -2.15$) between \textit{G1} ($Mean = 4.17$, $SD = 1.03$) and \textit{G2} ($Mean = 5.08$, $SD = 1.08$), which indicates that {\name} better helped improve voice modulation skills.}

%

%% file: discussion.tex

%% file: conclusion.tex
\section{Discussions and Limitations}
\textcolor{black}{
{\name} is designed to provide novice speakers with evidence-based training of voice modulation skills. 
Our in-depth expert interviews and user study provide support for the usefulness, effectiveness, and usability of {\name} in facilitating the training of voice modulation skills. However, there are still several key aspects that need further discussions.
}

\textcolor{black}{\textbf{Lessons learned} We summarize the important lessons learned from our system implementation, and evaluation studies. 1) \textit{Design a progressive learning process for skill acquisition.} 
During our design process, experts pointed out that it is challenging for novice speakers to apply all kinds of voice modulation skills to one sentence and to master new skills in one try.
To ease the training process and to improve learning efficiency,
our system promotes deliberate and iterative voice modulation practice on words of interests.
During our user study, participants acknowledged the design of {\pv} as helping them focus on specific parts of the sentence and gradually improving their skills by highlighting issues in their previous practices. Thus, we expect that the system should develop strategies of breaking down the overall training goal into small tasks and giving users step-by-step instructions on the tasks.
2) \textit{Turn practice into a game.} During the user study, we observed that several participants spent extra time interacting with voice curves in the practice view. 
They tried all the example sentences with different recommended modulation skills, and reported that interacting with real-time feedback was like playing a game.
This indicates that adding interesting designs in the training system can increase user engagement, benefiting successful learning of skills. 
3) \textit{Provide flexible and personalized training.}
In the expert interviews, we found that coaches had 
very different preferences for voice modulation skills,
which may lead to a biased training of specific types of voice modulation in the traditional methods of public speaking training, and a failure to meet the needs of speakers from different backgrounds.
These subjective biases provide support for the importance of a flexible and personalized training.
}

\textcolor{black}{
\textbf{Effectiveness and usability evaluations} Our current evaluations consist of in-depth interviews with
domain experts and user studies with
university students, which can provide support for the evaluation of the effectiveness and usability of {\name}.
The current system will be deployed to the public speaking training platforms of our industry collaborator.
With more participants from diverse background, it will further evaluate and verify the effectiveness and usability of {\name}.
}

\textbf{Technical limitations of {\name}}
First, we use TED talks as the benchmark dataset. Though 2,623 high-quality speeches are included, they may still do not cover all the ``good'' speeches in different domains. For example, the desirable voice modulation skills for an academic talk can be different from that for a business talk, but there are not many academic talks in the TED dataset.
Second, {\name} currently focuses on recommending ``good'' voice modulation examples and how to help speakers to learn from ``bad'' examples is not explored, which, as mentioned by the coaches in our expert interviews, may be also beneficial to the voice modulation training.
Third, our current recommendation of voice modulation examples is mainly based on the similarity of sentence structures, \textcolor{black}{which does not consider the preferences of different users in different speaking scenarios.}

\section{Conclusion and Future work}
In this paper, we \textcolor{black}{present} {\name}, an interactive evidence-based training system for voice modulation skills in public speaking. 
By working closely with professional communication coaches from our industry collaboration company in the past eight months, we have identified two of the most important major requirements of effective voice modulation training: \textit{concrete and personalized guidelines} and \textit{on-the-fly} feedback. 
Accordingly, we analyzed 2,623 high-quality TED speeches and recommend voice modulation examples to users based on the sentence structure similarity between the voice input and the TED speech segments, providing users with evidence-based hints on improvements of their vocal skills.
{\name} further enables quantitative and immediate feedback, through comparing the volume, pitch, and speed of users' voice input with their prior practice, to guide their further improvement on voice modulation skills.
Our semi-structured expert interviews and \textcolor{black}{user study with university students}
provide support for the good usability and effectiveness of {\name} in helping \textcolor{black}{novice speakers} with the training of voice modulation skills.


In future work, we would like to extend the current benchmark dataset by including the speeches in different domains (e.g., academic talks, public campaigns), 
and further 
improve the applicability of {\name}.
It would also be interesting to collect ``bad'' examples of voice modulation and improve the current system by showing negative examples as well to users, informing them of the voice modulation mistakes they should avoid.
\textcolor{black}{Furthermore, we plan to invite more participants with more diverse backgrounds, to further validate the usability and effectiveness of {\name} in helping novice speakers with evidence-based training of voice modulation skills.}
 

%% file: acks.tex
\section{acknowledgements}
The authors would like to thank 
the experts and participants for their help in the project,
as well as
the anonymous reviewers for their valuable comments. This project is funded by a grant from ITF UICP (Project No. UIT/142).